# From Network Sharing to Multi-tenancy: The 5G Network Slice Broker

Konstantinos Samdanis, Xavier Costa-Perez, Vincenzo Sciancalepore
NEC Europe Ltd, Germany
samdanis@neclab.eu, xavier.costa@neclab.eu, vincenzo.sciancalepore@neclab.eu

**Abstract**
The ever-increasing traffic demand is pushing network operators to find new cost-efficient solutions towards the deployment of future 5G mobile networks. The network sharing paradigm was explored in the past and partially deployed. Nowadays, advanced mobile network multi-tenancy approaches are increasingly gaining momentum paving the way towards further decreasing Capital Expenditures and Operational Expenditures (CAPEX/OPEX) costs, while enabling new business opportunities. This paper provides an overview of the 3GPP standard evolution from network sharing principles, mechanisms and architectures to future on-demand multi-tenant systems. In particular, it introduces the concept of the 5G Network Slice Broker in 5G systems, which enables mobile virtual network operators, over-the-top providers and industry vertical market players to request and lease resources from infrastructure providers dynamically via signaling means. Finally, it reviews the latest standardization efforts considering remaining open issues for enabling advanced network slicing solutions taking into account the allocation of virtualized network functions based on ETSI NFV, the introduction of shared network functions and flexible service chaining.

## 1. Introduction

Network sharing has evolved from a novel concept a few years back with the arrival of 3G networks to a fundamental feature of the emerging 5G systems. Mobile operators are facing tremendous traffic increases with the introduction of smartphones and tablets, especially due to content rich multimedia and cloud applications, and the upcoming vertical market services in automotive, e-health, etc. [1]. The challenge for mobile operators is to accommodate such traffic volumes without significantly increasing operational and infrastructure costs. The trend toward network densification for increasing network capacity and the practice of overprovisioning to accommodate peak demands including future traffic volumes adds further burden into the operational complexity and cost, diminishing the Return of Investment (RoI).

Indeed, 50% of the radio access sites carry traffic that yields less than 10% of the revenue [3]. Consequently, there is a need for mobile operators to exploit new revenue sources and break the traditional business model of a single network infrastructure ownership. Network sharing can recover up to 20% of operational costs for a typical European mobile network operator and can at least half the infrastructure cost of passive Radio Access Network (RAN) components, which make up to 50% of the total network cost [2]. An overview of network sharing CAPEX/OPEX saving on different parts of the network is depicted in Fig.1, considering the RAN, backhaul and core networks[1].

---

[1] Source of CAPEX/OPEX savings: Mobile Network infrastructure sharing – Industry Overview & Coleago's Approach, Coleago Consulting, Feb. 2015.

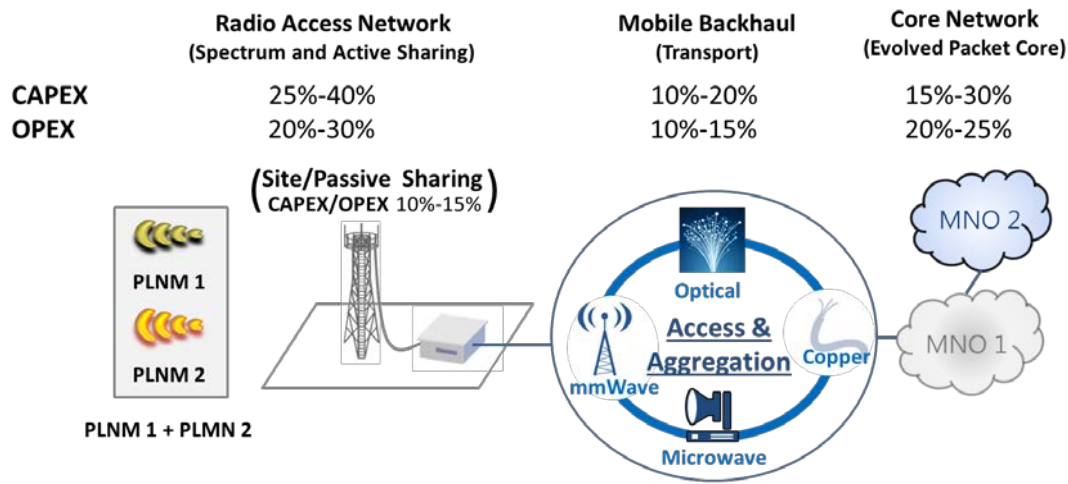

**Figure 1: Overview of CAPEX/OPEX saving of network sharing**

Mobile operators can share network infrastructures accelerating network roll-outs and offer services to customers with reduced costs. In urban areas network sharing can help avoiding complex and lengthy processes for site acquisition due to regulation issues, especially in highly populated regions where dense deployments restrict the available space, while for rural areas sharing can reduce the network investment payback period. Two different roles can be defined for network sharing solutions:

(i) <u>Infrastructure provider (InP)</u> responsible for the physical network deployment and maintenance. Mobile Network Operators (MNOs) or third parties that interact with other "players" but not with end users directly can take the InP role.
(ii) <u>Mobile Virtual Network Operator (MVNO)</u> lacks network infrastructure or has limited capacity and/or coverage, and leases resources from an existing InP.

Future multi-tenant systems are envisioned to expand the aforementioned roles to also include:

(iii) <u>Over-The-Top (OTT) service providers</u> operating on top of a network infrastructure belonging to an MNO and based on a pre-defined Service Level Agreement (SLA) set of requirements.
(iv) <u>Vertical industries</u> exploiting an MNO network infrastructure for services complementary to the telecommunication industry.

In both cases the allocated network slices can be provided on a permanent basis or on-demand, i.e. opportunistically or periodically.

This paper provides an overview of the 3GPP standardization activities on network sharing, focusing on the business requirements, architectures and network management framework. In addition, it introduces the main enablers for realizing future flexible, on-demand multi-tenant networks. The main contribution of this paper is the analysis and design of a signaling-based, i.e. with no human intervention, on-demand multi-tenant network building on the top of the 3GPP network sharing management architecture. In the core of our proposed on-demand multi-tenant network architecture lies a logically centralized monitoring and control entity defined as *5G Network Slice Broker* providing admission control for incoming requests (placed by MVNOs, OTTs and Verticals) and resource assignment by means of an enhancement of the 3GPP network sharing management architecture interfaces and Service Exposure Capability Function (SECF).

The remaining of this paper is organized as follows. Section 2 presents the fundamental network sharing scenarios and their corresponding business requirements. Section 3 describes the 3GPP network sharing standardization efforts and architectures. Section 4 introduces the network sharing management architecture for supporting mobile virtual operators along with the ongoing enhancements for supporting vertical industries. Section 5 analyses the proposed *5G Network Slice Broker* architecture. Section 6 presents the 3GPP efforts on evolved network slicing towards the realization of full multi-tenancy considering open standardization issues. Finally, Section 7 provides the conclusions.

## 2. Network Sharing Scenarios and Business Requirements

The adoption of network sharing and multi-tenancy from the business perspective aims to address different strategic and commercial targets for each participant player. For InPs, network sharing results in additional revenue sources and thus better return on CAPEX/OPEX investments. MVNOs exploit network sharing as a mean to enhance service provisioning in regions with low or no network footprint, where the payback period is estimated greater than the expected business targets.

In general, the adoption of network sharing in mature markets concentrates on increasing RoI and capacity enhancement. In developing markets network sharing usually focuses on coverage expansion. A significant aspect that influences MNOs' decision of whether enabling network sharing is beneficial for their business relies on the purpose of sharing and on the risk of reducing their competitive advantage. For instance allowing coverage enhancements of their competitors is sensitive for emerging mobile markets where coverage is a significant service attribute, but it becomes more relaxed in cases where QoS provision and service innovation is the key business differentiator. The 3GPP Services Work Group SA1 specified five main business scenarios for network sharing in [5] summarized here:

- **Multiple core networks sharing a common RAN**: An early scenario considered in 3GPP Release 99, where operators share RAN elements, but not the spectrum. In this case operators connect directly to their own dedicated carrier layer in the shared Radio Network Controller (RNC) in the shared RAN.
- **Operator collaboration to enhance coverage**: In this scenario two or more operators with individual frequency licenses and respective RANs that cover different parts of a country, provide together coverage for the entire country.
- **Sharing coverage on specific regions**: In this scenario one operator provides shared coverage in a specific geographical area, with other operators allowed to use it for their subscribers. Outside such area, coverage is provided by each operator independently.
- **Common spectrum sharing**: This scenario corresponds to common spectrum RAN sharing considering the following two variations: i) One operator has a frequency license and shares the spectrum with other operators, ii) A number of operators decide to pool their allocated spectrum and share the total.
- **Multiple RANs share a common core network**: In this scenario multiple RANs share a common core network, with each RAN belonging to different network operator.

Challenging the traditional mobile communication paradigm by considering the evolution towards multi-tenancy on an open regulation environment provides the opportunity for commercial and operational separation of the mobile infrastructure from service layers. In this way they can evolve independently according to different business needs and performance characteristics.

## 3. Early Network Sharing Standardization and Architectures

In the early GSM and UMTS stage, network sharing support was not included; the mobile network design was concentrating on a single MNO. 3GPP Rel.99 introduced in the UMTS network the first generation of network sharing concentrating on simple solutions in terms of the commercial exploitation, with passive sharing and network roaming being the two main pillars. Passive sharing is defined as the sharing of site locations or physical supporting infrastructure of radio equipment, such as masts. Site sharing allows mobile operators to share space and optionally share certain support equipment such as shelters or power supply, but with separate installations of masts, antennas, cabinets and backhaul equipment. Such approaches did not gain significant interest from the industry until the early 2000s.

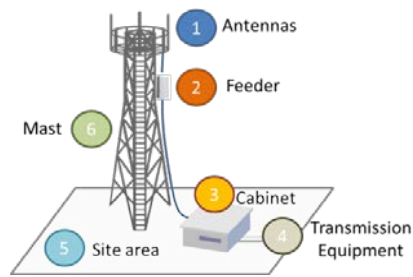

Figure 2: Passive network sharing

A step further was accomplished with mast sharing, where mobile operators can co-locate their sites and even share the antenna frame, but still install their own radio equipment, maintaining separate coverage. An overview of passive network sharing is illustrated in Figure 2 highlighting the main components. As for network roaming, certain mobile subscribers can use the network of other MNOs based on contractual agreements without imposing any particular network sharing requirements, so in that sense it is not exactly a form of infrastructure sharing. With 3GPP Rel-6 (UMTS), Rel-8 (LTE) and Rel-10 (LTE-A), new requirements were needed to shed the light on the potential of network sharing.

Active RAN sharing followed the first generation of network sharing, which focused on sharing access network equipment including base stations, antennas and even mobile backhaul equipment. In active RAN sharing MNOs can pool spectrum resources, which are shared alongside other RAN equipment based on fixed, contractual agreements. The interest in sharing the resources dynamically introduced new requirements, beyond the original RAN sharing concepts, where MNOs share core transmission equipment, billing platforms and core network equipment.

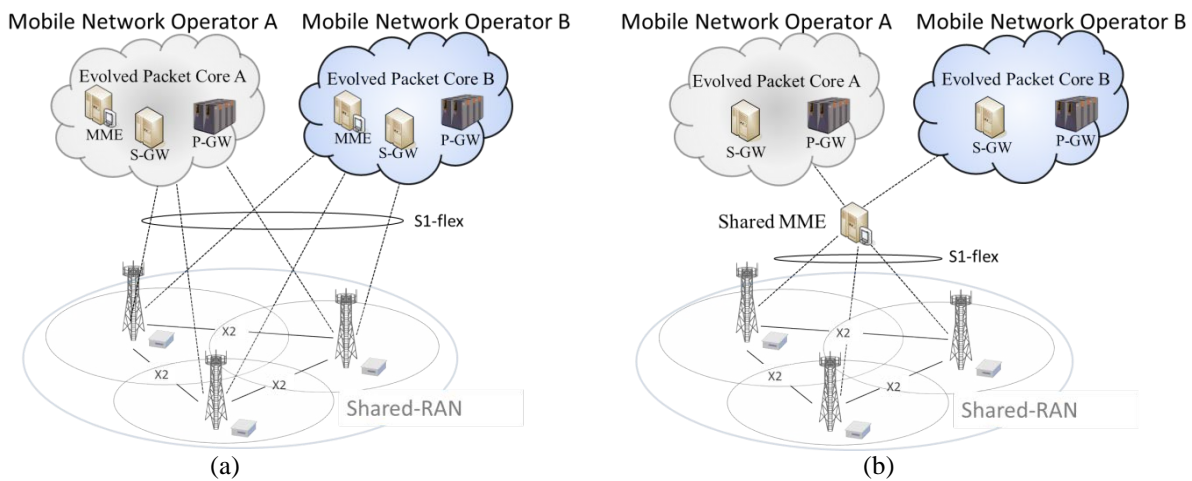

Figure 3: 3GPP architectures for network sharing (a) MOCN and (b) GWCN.

3GPP specified two distinct active RAN sharing architectures as illustrated in Figure 3 in the Architecture Working Group SA2 in Rel.11 – Rel.12 as documented in [4]:

- **Multi-Operator Core Network (MOCN),** where each operator has its own Enhanced Packet Core (EPC) providing a strict separation amongst the core network and RAN. Shared base stations, i.e. evolved Node Bs (eNBs), are connected to core network elements of each different operator, i.e. Mobility Management Entity (MME) and Serving/Packet-Gateway (S/P-GW), using a separate S1 interface. This enables customization, for example allowing load balancing policies to be provided within each operator's core network. MOCN offers benefits regarding service differentiation and interworking with legacy networks.
- **Gateway Core Network (GWCN),** where operators share additionally the MME; an approach that further enables cost savings compared to MOCN, but at the price of reduced flexibility, i.e. restricting mobility for inter-Radio Access Technology (RAT) scenarios and circuit switching fallback for voice traffic.

In general, MOCN requires a higher investment but is considered to be more flexible, addressing easier conventional MNOs' needs. The User Equipment (UE) behavior in both MOCN and GWCN is identical with resource sharing being transparent. In both cases, UEs can distinguish up to six different MNOs that share the RAN infrastructure based on broadcast information, i.e. Public Land Mobile Network (PLMN)-id, and can signal to obtain connectivity or perform a handover irrespective of the underlying RAN sharing arrangement. Specifically, the S1 interface supports the exchange of PLMN-ids between eNBs and MMEs in order to assist the selection of the corresponding core network, as documented in TS 36.413, while the X2 interface supports a similar PLMN-id exchange among neighboring eNBs for handover purposes, as per TS 36.423. Regarding broadcasting, the Uu interface supports the PLMN-ids enabling the UEs to perform the network selection as specified in TS 36.331.

## 4. Incorporating Virtual Operators & Verticals in 3GPP Networks

The 3GPP Telecom Management Working Group SA5 has extended the legacy network management architecture to accommodate network sharing based on long term contractual agreements [6]. Such network sharing paradigm considers that an InP, referred to as *Master Operator* in the 3GPP terminology, facilitates resource sharing to *Participant* MVNOs or otherwise *Sharing Operators* through the InP network manager system, using the Type 5 interface.

Type 5 interface is established upon an agreement between MNOs to provide connectivity among the network manager systems across different organizations, e.g. for roaming purposes. The Master Operator can then forward performance monitoring information through the network manager system, referred in 3GPP as Master Operator-Network Manager (MO-NM), to the participant Sharing Operator-Network Manager (SO-NM). For monitoring and configuration operations on network elements the MO-NM can use:

(i) **Type 2 interface or Itf-N** between the MO-NM and network element manager. In LTE the element manager is co-located at the eNB, while in UTRAN it is located on the Master Operator-Sharing RAN Domain Manager at the RNC. This interface is used for performance monitoring, reporting and control of network elements to the network manager system.
(ii) **Type 1 interface or Itf-B** between the Master Operator-Shared RAN Domain Manager and a NodeB. Typically, the Master Operator-Shared RAN Domain Manager serves a number of Shared RAN NodeBs. This interface is also used for network management purposes.

Vertical industries and OTT providers, which do not own a network infrastructure, need to interact with InPs to request network resources and to negotiate SLAs, a process that is achieved by allowing exposure of the 3GPP service capabilities to third parties. In this way operators are no longer merely suppliers of communication services, but business enablers. The 3GPP Service Capability Exposure Function (SCEF) [7] located at the operator trust domain provides a mean to securely expose selected service capabilities via network Application Programming Interfaces (APIs). The SCEF abstracts service capabilities related to the communication type, network elements, policy control and network resource allocation from the underlying 3GPP network. Effectively, such service capability abstraction can also assist third parties to issue a network resource request towards an InP. The SCEF plays the role of the mediator between the third party and the 3GPP InP facilitating the following operations:

(i) Authentication/authorization and secure access of third parties to the 3GPP network ensuring that the InP is under control of the exposed services,
(ii) Charging based on offered service and quality provision,

(iii) QoS provision and SLA monitoring, allowing third parties to request and set service priorities in a dynamic manner,
(iv) Provision of user context information, e.g., real-time user location, user connection properties, average data rate, etc., and network status changes to third parties,
(v) Admission control regarding predictable communication patterns, e.g. considering the time window and traffic volume, pre-schedule communication timing, etc.

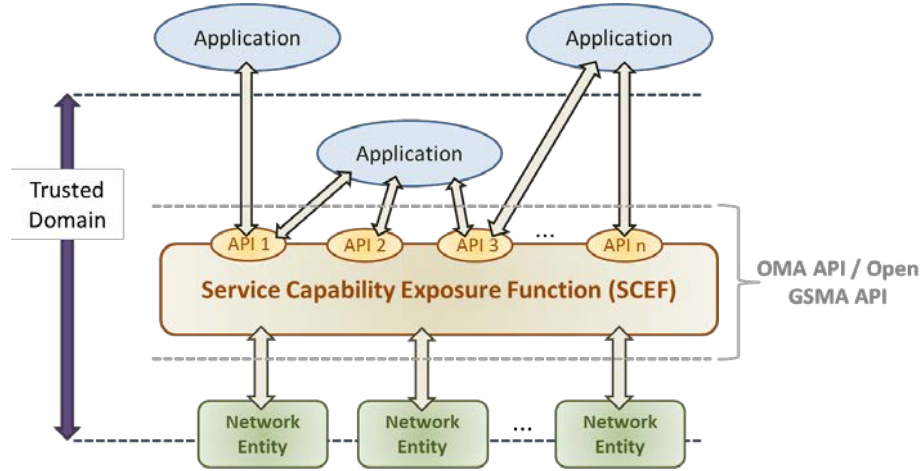

**Figure 4: Service Capability Exposure Function (SCEF) architecture**.

Such operations support the allocation of network resources with customized capabilities considering the developer's or third party's business requirements, SLA policy and service adaptation. Effectively, this provides the opportunity for network programmability allowing third parties to efficiently use the retrieved service capability information and optimally exploit the available network resources. Figure 4 illustrates the SCEF architecture showing how third party's applications can associate with different APIs receiving a customized set of service capabilities. The network elements and interfaces within the trusted domain are under the control of the InP, with the SCEF exposing the capabilities of network entities to the application layer.

## 5. 5G Network Slice Broker - Architecture

To enhance the existent RAN sharing flexibility, the authors of this paper introduced[2] in the 3GPP Services Working Group SA1 the concept of the *on-demand capacity broker* [5]. Differently to SCEF, which exposes service capabilities, the *on-demand capacity broker* facilitates on-the-fly resource allocation by allowing a host RAN provider, i.e. InP, to allocate via signaling means an indicated portion of network capacity for a particular time period to an MVNO, OTT provider or vertical market player.

In this paper, we build on top of the 3GPP SA5 network sharing management architecture, introducing a novel concept of capacity broker with a more generic objective, in order to address dynamic resource sharing scenarios by establishing network slices. A network slice refers to an isolated amount of network capacity customized to suit best specific service requirements. The proposed capacity broker, namely *5G Network Slice Broker*, can facilitate on-demand resource allocation and perform admission control based on traffic monitoring and forecasting including mobility based-on a global network view. In addition, it configures RAN schedulers to either follow a two-layer paradigm, with the higher layer operating an inter-slice resource allocation and the lower one enabling tenants to customize scheduling on the allocated spectrum in isolation or configure policies to enable the selection of resource blocks from a shared pooled spectrum, taking into account the service SLA and the size of the network slice across the core network.

---
[2] 3GPP S1-122194, On-demand Capacity Brokering, TSG-SA WG1 Meeting #59, NEC/Sprint, Aug. 2012.

To accomplish this task, we propose to co-locate the *5G Network Slice Broker* at the MO-NM, which monitors and controls the shared RAN, while interacting with the Sharing Operator Network Manager (SO-NM). The *5G Network Slice Broker* can gain in this way access to network monitoring measurements such as load and various Key Performance Indicators (KPIs), e.g. mobility optimization, failures, SLA violations, etc. as well as obtain network infrastructure capabilities information. In addition, it can receive on-demand network resource requests from MVNOs, via the Type 5 interface, for allocating network slices based on SLAs. The *5G Network Slice Broker* upon performing the corresponding admission control decisions, it can then take advantage of the existing MO-NM interfaces, i.e. Itf-N and Itf-B, to configure the desired network slice on specific RAN network elements.

Besides MVNO requests, the *5G Network Slice Broker* can also handle requests with a specified SLA from a range of vertical industries and OTT providers, through a close interaction with the SCEF or exploiting the co-location of the SCEF at the MO-NM. The interface of verticals or OTT providers towards the SCEF is under discussion, with 3GPP adopting APIs developed in other standardization bodies, e.g. the Open Mobile Alliance (OMA) API focusing on sensor/machine type applications and the GSM Association (GSMA) Open API designed for application providers. In this way the SCEF and the corresponding API is not only exposing information about devices, but can also provide control to vertical industries and OTT providers through the *5G Network Slice Broker* and the MO-NM interfaces, i.e. Itf-N and Itf-B, to allocate the desired SLAs.

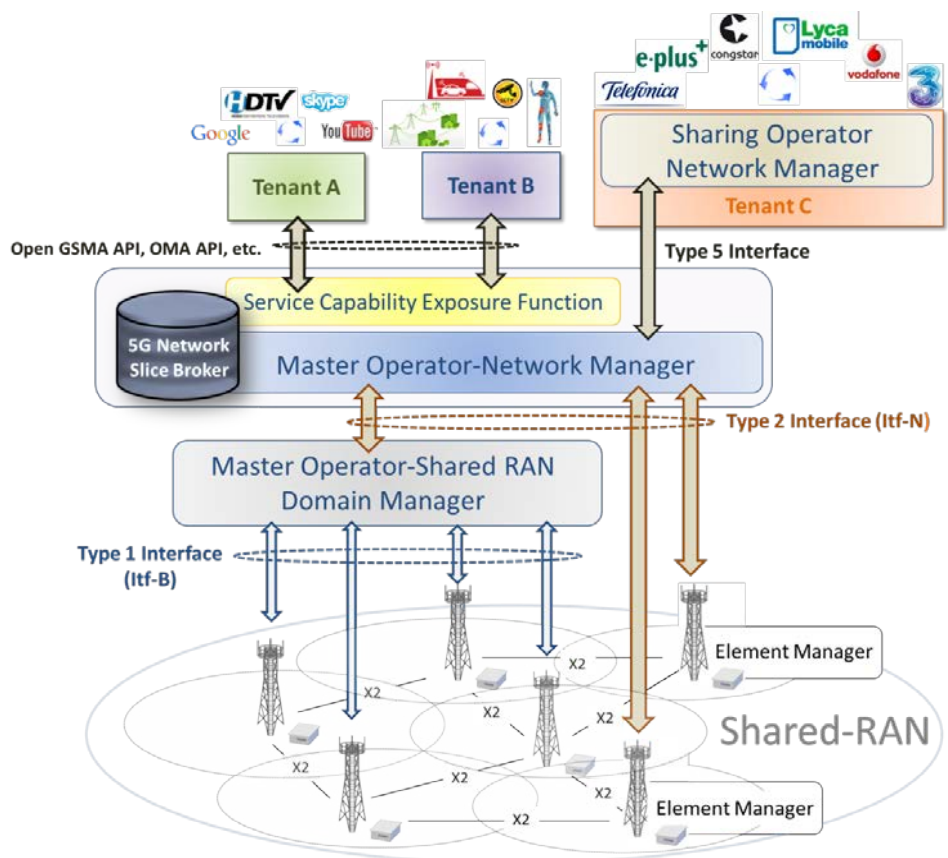

**Figure 5: 5G Network slice broker- Management architecture**

An overview of the proposed 3GPP-compliant network slice broker management architecture is illustrated in Figure 5, showing the *5G Network Slice Broker* and the SCEF co-located at the MO-NM. The SCEF provides access to OTT providers, e.g. video, voice and social applications, etc., as well as to vertical industries, e.g. electricity utility, automotive, e-health, etc. A direct connection through the network managers, i.e. MO-NM and SO-NM, enables various tenants to easily access RAN resources. Hence, the *5G Network Slice Broker* acts as mediator, mapping the SLA requests of multiple tenants with the physical network resources through the interfaces provided by the MO-NM.

Interestingly, the proposed *5G Network Slice Broker* management architecture supports on-demand resource allocation operations. This can be achieved by enhancing the existing interfaces. In particular, enhancements should differentiate tenants in order to handle the corresponding data traffic and provide performance monitoring information towards each participant operator through Type 5, Itf-N and Itf-B interfaces. To enable this, a tenant identifier, e.g. PLMN-id, can be included in each data packet corresponding to different slices as well as in performance measurement reports to enable the MO-NM to provide feedback towards the corresponding SO-NM. Such performance feedback should involve only the allocated slice resources for privacy and competition reasons. For supporting verticals and OTT providers the Itf-N ad Itf-B should also be enhanced to distinct these types of tenants by introducing a corresponding service identifier to each data packet and performance monitoring report.

The Type 5 interface as well as the vertical industries/OTT provider APIs should be extended to accommodate on-demand network slice requests with a particular SLA and timing requirements. The Itf-N and Itf-B interfaces should also be extended to carry out the configuration of network slices by introducing a new type of signaling considering MVNOs and vertical industries/OTT providers. Such interface enhancement and signaling should contain a set of additional information including:

(i) the amount of resources allocated to a network slice, e.g. physical resources or data rate
(ii) timing, e.g. starting time, duration or periodicity of a request, time window
(iii) the type of resources and QoS, e.g., the radio/core bearer type, prioritization, delay, jitter, loss
(iv) the size of file to be downloaded or data to be communicated to particular device/user or application server
(v) service related information, e.g. mobility (stationary, low, medium, high), data offloading policies, service disruption tolerance

Besides the service characteristics of a network slice, the set of cells which should accommodate the network slice request is an additional parameter that can be considered. Effectively, such a parameter can be either explicitly provided by the MVNO via Type 5 interface or it can be determined by the InP considering the location of the corresponding devices/users in combination with tailored service information provided by the *5G Network Slice Broker*. The set of cells that need to accommodate a network slice should be communicated via the Itf-N interface towards the Master Operator – Shared RAN Domain Manager, which in turn would configure the appropriate cells using the Itf-B interface.

# 6. The Network Slicing Road towards Full Multi-tenancy

The evolution of network sharing towards full multi-tenancy relies on virtualization mechanisms and software-based capabilities which are progressively introduced into 3GPP networks, influencing its standardization roadmap. These capabilities enhance the notion of *network slicing* for supporting particular communication services. Such emerging network slicing will be realized by allocating not only network capacity, but also Virtual Network Functions (VNFs), computing resources, per slice tailored control/user-plane splits, shared network functions across different slices and RAT settings as described by NGMN in [8]. Network slicing can further be enriched accommodating particular applications, which can be located at the network edge to improve end-users' performance. 3GPP SA1 emphasizes the support for vertical industries via network slicing in Release 14 considering terminal operations and configuration management via suitable APIs. The main attributes to realize the aforementioned network slicing extensions in 3GPP towards full multi-tenancy are:

- **Network Function Virtualization:** 3GPP has adopted ETSI NFV MANO [9] shedding light on the potential impact of virtualized networks on the existing 3GPP SA5 network management architecture [10], considering partially and entirely VNFs with respect to macro-base stations and core network elements. The objective is to identify requirements, interfaces and procedures, which can be re-used or extended for managing virtualized networks. In Release 14 3GPP has introduced a specification on architecture requirements for virtualized network management [11], considering complementary specifications on configuration, fault performance and life-cycle management. An equivalent study focusing on small cells and on the adoption of flexible Centralized-RAN has been performed at the Small Cell Forum [12].

- **Dedicated Core Networks (DCNs):** In an effort to support devices/customers with different service characteristics including vertical market players, 3GPP SA2, introduced in Release 13 the support of separate DCNs [13], with different operation features, traffic characteristics, policies, etc. Each DCN is assigned to serve different types of users based on subscription information, assuring resource isolation and independent scaling, offering specific services and network functions including RATs. Effectively, the *5G Network Slice Broker* may allocate a collection of shared network resources and VNFs among particular slices that fulfill the requirements of certain communication services.

- **User/Control-plane Separation and Service Chaining**: 3GPP has initiated a study on user/control-plane separation in TR 23.714 analyzing potential architecture enhancements for core network elements, e.g. Packet Data Network Gateway (P-GW), Traffic Detection Function (TDF), etc., to further enable flexibility for the network deployment and operation, enabling a unified network management across different RATs. An equivalent process focusing on services, e.g. firewall, Deep Packet Inspection (DPI), etc., should also be considered when establishing network slices. 3GPP has performed in TR 23.718 (Rel.13) a study on flexible mobile service steering focusing on policy provision and on instantiating dynamic services in SGi-LAN, a service-oriented network connected to the P-GW. A comprehensive study on flexible service chaining on mobile networks, considering a range of different service chaining mechanisms is provided in [14].

- **Mobile Edge Computing (MEC)**: Many evolving 5G services are envisioned to be offered closer to the user at the network edge in order to enhance latency and in general end-user perceived performance, e.g. adopting the ETSI MEC paradigm [15]. Hence, flexible service chaining should also be enhanced to establish dynamic services considering edge network locations and potentially be combined with VNFs, in order to enable a joint optimization of services and networking. Edge server locations can also be exploited for storage, computation and dynamic service creation by verticals/OTT providers, introducing in this way another multi-tenancy dimension.

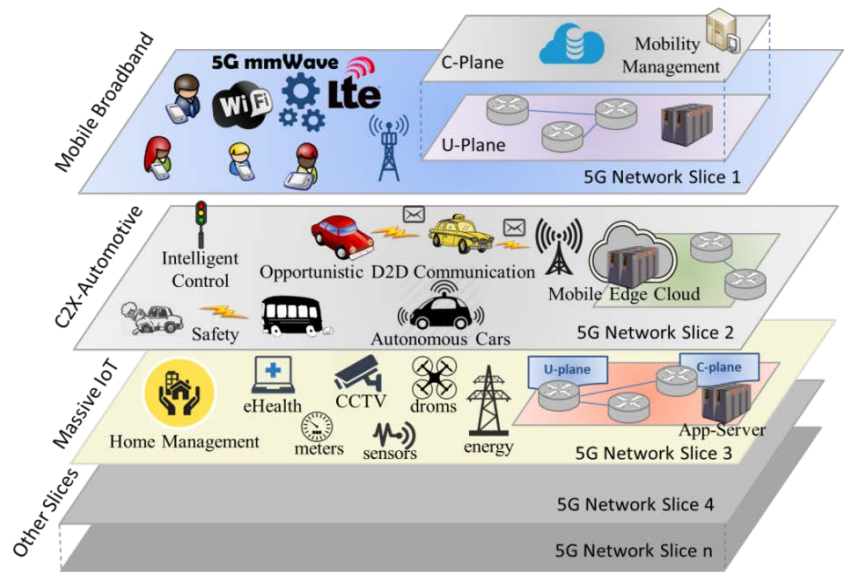

Figure 6: 5G network slices structure

Figure 6 illustrates an example of different network slices operating on the same infrastructure: i) a network slice that accommodates mobile broadband services, ii) an automotive network slice wherein latency and reliability are critical parameters and iii) a massive Internet of Things (IoT) network slice where scalability is essential for handling efficiently huge amounts of small data. To accommodate strict latency goals and scalability, network functions can be instantiated at the edge cloud as necessary, optimizing radio and core networks with respect to particular services. Different RATs should be associated with distinct types of network slices, since they can serve best the requirements of particular services.

## 7. Conclusion

This paper reviewed the path from network sharing towards multi-tenancy describing business requirements and standardization efforts with a focus on 3GPP. In particular, it analyzed the evolutionary path from early passive sharing to on-demand multi-tenant networks considering: i) 3GPP network sharing architectures, ii) network management extensions for supporting mobile virtual operators and iii) the service exposure capability function that allows vertical market players to gather information about mobile network resources. The notion of the *5G Network Slice Broker* has been introduced, which resides inside the infrastructure provider, detailing the required interfaces and functional enhancements for supporting on-demand multi-tenant mobile networks based on the latest 3GPP network sharing management architectures. Finally, our work provided an overview of the 3GPP Rel.14 standardization efforts related to multi-service support and network virtualization as well as other relevant standardization efforts outside 3GPP addressing how to enrich a 5G network slice by flexibly provisioning virtualized network functions and services.

## Acknowledgement

This work has been performed in the framework of the H2020-ICT-2014-2 project 5G NORMA. The authors would like to acknowledge the contributions of their colleagues. This information reflects the consortium's view, but the consortium is not liable for any use that may be made of any of the information contained therein.

## References

[1] GSMA, The Mobile Economy 2016.

[2] GSMA, Mobile Infrastructure Sharing, Sep 2012.

[3] K. Larsen, "Network Sharing Fundamentals," Technology Business, Jul. 2012.

[4] 3GPP TS 23.251, Network Sharing; Architecture and Functional Description, Rel.12, Mar. 2015.

[5] 3GPP TR 22.852, Study on Radio Access Network (RAN) Sharing enhancements, Rel.13, Sep. 2014.

[6] 3GPP TS 32.130, Telecommunication management; Network Sharing; Concepts and requirements, Rel.12, Jan. 2016.

[7] 3GPP TR 23.708, Architecture enhancement for Service Capability Exposure, Rel.13, Jun. 2015.

[8] NGMN Alliance, NGMN 5G White paper, version 1, Feb 2015.

[9] ETSI GS NFV-002 Architectural Framework, v1.2.1, Dec. 2014.

[10] 3GPP TR 32.842, Telecommunication management; Study on network management of virtualized networks, Rel.13, Dec. 2015.

[11] 3GPP TS 28.500, Management Concept, Architecture and Requirements for Mobile Network that include Virtualized Network Functions, Rel.14, Jan. 2016.

[12] Small Cell Forum, Network Aspects of Virtualized Small Cells, Release 5.1, Document 161.05.1.01, Jun. 2015.

[13] 3GPP TS 23.401, General Packet Radio Service (GPRS) enhancements for Evolved Universal Terrestrial Radio Access Network (E-UTRAN) access, Rel.13, Dec. 2015.

[14] W. Haeffner, J. Napper, M. Stiemerling, D. Lopez, J. Uttaro, Service Function Chaining Use cases in Mobile Networks, IETF Draft, Version 5, Oct. 2015.

[15] ETSI, Mobile-Edge Computing – Introductory White paper, Sep. 2014.

## Biographies

**Konstantinos Samdanis (samdanis@neclab.eu)** is a Senior Researcher and Backhaul/Broadband standardization specialist with NEC Europe. He is involved in research for 5G architectures participating in 5G-NORMA and is active in BBF on 5G and network virtualization. Konstantinos has provided a number of tutorials in IEEE conferences for Green Communications and is the editor of the Green Communications: Principles, Concepts and Practice book from Wiley. He received his Ph.D. and M.Sc. degrees from Kings College London.

**Xavier Costa-Pérez (xavier.costa@neclab.eu)** is Head of 5G Networks R&D at NEC Laboratories Europe, where he manages several projects focused on 5G mobile core, backhaul/fronthaul and access networks. His team contributes to NEC projects for products roadmap evolution as well as to European Commission R&D collaborative projects and has received several R&D Awards for successful technology transfers. In addition, the 5G team contributes to related standardization bodies: 3GPP, BBF, ETSI NFV, ETSI MEC and IETF.

**Vincenzo Sciancalepore (vincenzo.sciancalepore@neclab.eu)** (S'11-M'15) received his M.Sc. degree in Telecommunications Engineering and Telematics Engineering in 2011 and 2012, respectively, whereas in 2015, he received a double Ph.D. degree. From 2011 to 2015 he was Research Assistant at IMDEA Networks, focusing on inter-cell coordinated scheduling for LTE-Advanced networks and device-to-device communication. Currently, he is a Research Scientist at NEC Laboratories Europe in Heidelberg, focusing his activity on network virtualization and network slicing challenges.